\documentclass{article}
\usepackage[preprint]{neurips_2026}
\usepackage[utf8]{inputenc} % allow utf-8 input
\usepackage[T1]{fontenc}    % use 8-bit T1 fonts
\usepackage{hyperref}       % hyperlinks
\usepackage{url}            % simple URL typesetting
\usepackage{booktabs}       % professional-quality tables
\usepackage{amsfonts}       % blackboard math symbols
\usepackage{amsmath}        % math environments
\usepackage{nicefrac}       % compact symbols for 1/2, etc.
\usepackage{microtype}      % microtypography
\usepackage{cleveref}       % intelligent cross-referencing
\usepackage{xcolor}         % colors
\usepackage{graphicx}       % figures
\usepackage{subcaption}     % subfigures
\usepackage{algorithm}      % algorithm environment
\usepackage{algpseudocode}  % pseudocode
\usepackage{multirow}       % multirow table cells
\usepackage{enumitem}       % itemize controls
\usepackage{float}          % [H] force-here placement
\usepackage{adjustbox}

\title{NTILC: Neural Tool Invocation via Learned Compression}

\author{%
  Andrew Krikorian \\
  Department of Robotics\\
  University of Michigan\\
  Ann Arbor, MI 48104 \\
  \texttt{akrik@umich.edu} \\
  \And
  Yayuan Li \\
  Department of ECE \\
  University of Michigan\\
  Ann Arbor, MI 48104 \\
  \texttt{yayuanli@umich.edu} \\
  \And
  Jason Corso \\
  Department of Robotics\\
  University of Michigan\\
  Ann Arbor, MI 48104 \\
  \texttt{jjcorso@umich.edu} \\
}

\begin{document}

\maketitle
\begin{abstract}
Agentic tool-calling language models depend on large registries of callable APIs, functions, and local actions.
Placing full tool specifications directly in the prompt incurs a cost that scales linearly with the size of the tool registry, rapidly consuming the context budget. As the registry grows, this leads to higher latency and degrades selection accuracy, particularly due to interference from irrelevant tools.
We overcome these limitations by introducing NTILC, a neural tool selection and invocation framework that replaces in-context registry look-up with learned latent retrieval.
NTILC maps both user intent and tool specifications into a shared embedding space, enabling tool selection via external retrieval rather than in-context lookup. The language model is conditioned only on the selected tool schema, allowing for precise, constrained argument generation.
Central to our approach is a signature-aware composite objective, which augments semantic similarity with constraints derived from tool signatures (e.g., argument schema, type compatibility, and return types). By combining Circle Loss with a Functional Margin Loss, the model enforces separation between tools that are semantically similar but incompatible under their execution signatures.
We evaluate NTILC on public tool-selection and function-calling datasets and report context token usage, retrieval accuracy, and selection latency metrics.
Across these settings, NTILC reduces context window consumption by over 95\% and inference latency by up to 74\% compared to long-context ICT baselines.
\end{abstract}

\section{Introduction}
\label{sec:intro}

Language models have increasingly become the reasoning core of agents \cite{parisi2022talmtoolaugmentedlanguage, li2025teachinglanguagemodelsreason, Plaat_2025}. In this architectural framework, agents are defined as LLMs augmented with tool registries, operating through a modular pipeline comprising task planning, tool discovery, parameter generation, and response synthesis. Modern agents extend these capabilities by increasing their tool registries: web APIs, code interpreters, database queries, and shell commands, instead of encoding all information in model weights. While this integration has unlocked a new tier of agentic autonomy, it has also induced a practical bottleneck due to vastly increased token usage, causing latency issues and context rot.

The standard approach to tool-augmented inference \cite{Mialonetal2023, shtok2024augmentingincontextlearningllmsautomatic, ye2026incontextreinforcementlearningtool}, which we refer to as In-Context Tooling (ICT), works as follows: before the model sees the user's request, every available tool definition is inserted into the context window.
The model then receives a query from the user, reads the full registry, and generates a structured invocation.
This approach is simple and effective for small tool registries, but it imposes a token cost that scales linearly with the number of tools.
Across the datasets used in our evaluation, tool schemas contain approximately 120 tokens on average.
Therefore, a registry with $5,000$ tools would require roughly $600,000$ input tokens before the model sees any task-specific information.
Although a long-context model can technically accommodate such registries, the overhead causes increased inference costs and degrades accuracy \cite{liu2023lostmiddlelanguagemodels, hosseini2024efficientsolutionsintriguingfailure, wang2026intelligencedegradationlongcontextllms, Paulsen_2026}.

To address these limitations, we propose \textbf{NTILC} (Neural Tool Invocation via Learned Compression).
Rather than forcing the agent to read raw text schemas at every call, NTILC compresses the entire tool registry into an embedding space.
At inference time, the agent produces a natural language intent: a concise description of the action required to obtain a specific piece of information. This intent serves as a query, which is embedded and used to retrieve the most relevant tool via nearest-neighbor search in the shared embedding space.
This eliminates the tool registry's footprint from the context window entirely, reducing tool registry prompt tokens from $O(N)$ to $O(1)$ regardless of how many tools are registered.

A key challenge in tool selection is that tools frequently share nearly identical natural language descriptions yet have incompatible arguments. For example, \texttt{getWeather(zipcode: int)} and \texttt{getWeather(lat: float, lon: float)} describe the same concept but cannot be substituted for one another.
This causes models to select semantically relevant but functionally incompatible tools. We term this phenomenon \textit{semantic blur}.

Our contributions:
\begin{itemize}[leftmargin=*,nosep]
    \item We introduce \textbf{NTILC}, a framework that replaces linear-scaling in-context registry scanning with an external learned dispatch step, reducing registry prompt tokens from $O(N)$ to $0$ with respect to the number of registered tools.
    \item We propose a \textbf{Functional Margin (FM) Loss} that applies a repulsive force between tools whose descriptions are similar but whose executable signatures differ. Combined with Circle Loss for query-tool alignment, this composite objective produces an embedding space that is not only semantically coherent but also functionally compatible.
    \item We evaluate NTILC on \textbf{public tool-selection and function-calling datasets} including ToolBench, BFCL, API-Bank, MetaTool, and ToolEyes, and report token cost, latency, and accuracy under increasing registry size.
\end{itemize}

\begin{figure}
    \centering
    \includegraphics[width=1\linewidth]{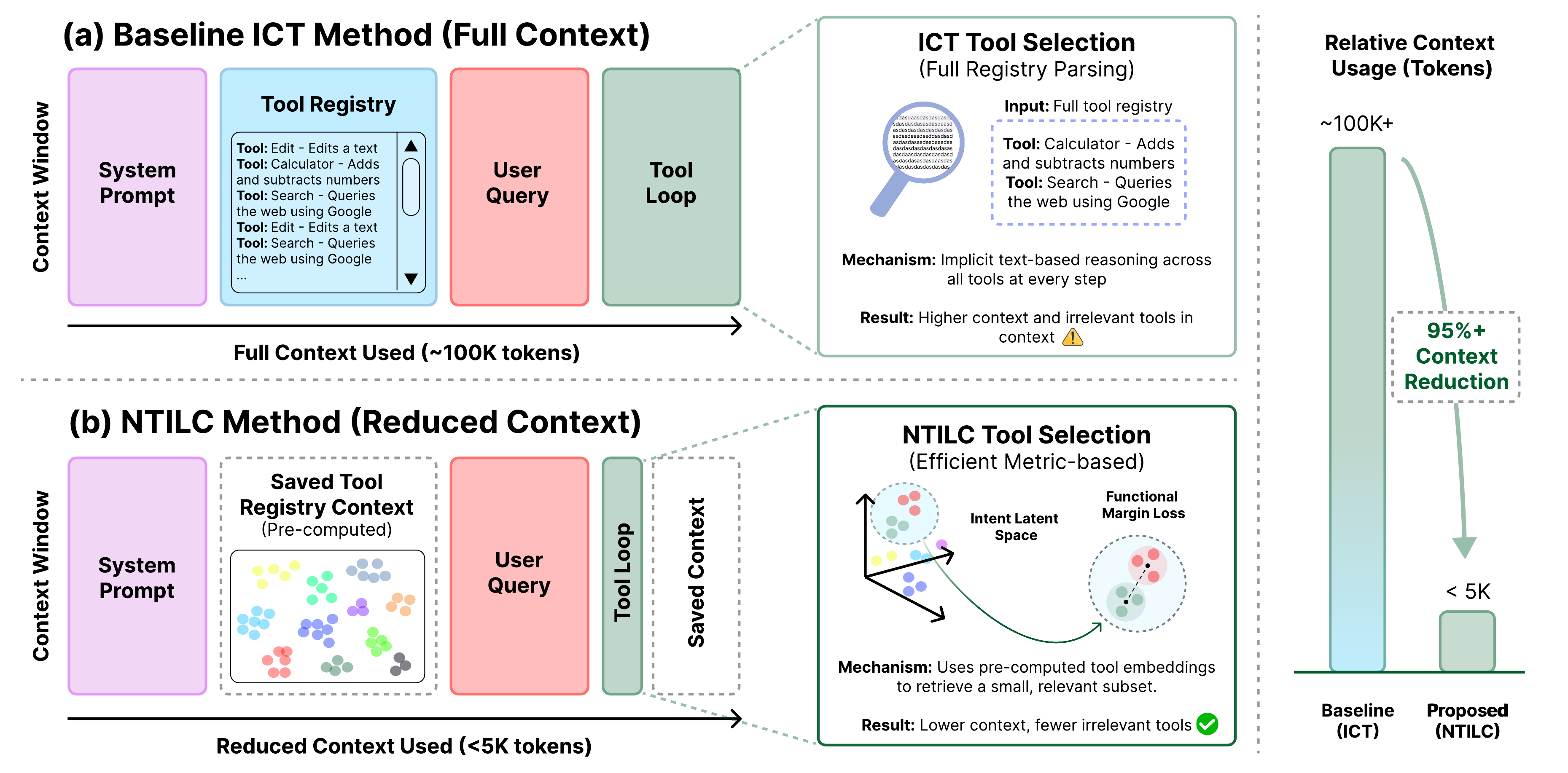}
    \caption{Relative context reduction achieved by NTILC compared to baseline ICT methods.}
    \label{fig:main}
\end{figure}

\section{Related Work}

The evolution of tool-augmented large language models has primarily focused on expanding the knowledge boundaries of models through external interfaces \cite{Mialonetal2023}.
We survey three lines of work that are most relevant to NTILC.

\paragraph{In-Context Tooling (ICT) and Prompt-Based Methods.}
Prompt-based tool-use systems provide tool descriptions directly in the system prompt \cite{Mialonetal2023}.
ICT is effective for small toolsets but suffers from a linear registry cost: as new tools are added, the registry consumes a growing share of the context window, increases prefill work, raises input-token costs, and exposes the model to more distractor tools \cite{Patiletal2023, Wangetal2024}.
NTILC addresses this by removing the full registry from the prompt.

\paragraph{Retrieval-Augmented Tool Selection.}
As tool libraries scale to hundreds or thousands of APIs, research has shifted toward retrieval-based architectures \cite{Qinetal2023}.
Typical implementations use sparse or dense retrieval to fetch a small set of relevant tool schemas into the context window dynamically \cite{Patiletal2023, Qinetal2023}.
This reduces the prompt from the full registry to top-$k$ candidate schemas, but it does not remove tool text from the LLM prompt and remains vulnerable to semantically plausible distractors \cite{Wangetal2024}.
Semantic retrieval often fails when tools share a topic but have incompatible functional signatures.
Unlike standard RAG-style tool selection, NTILC treats retrieval as the dispatch decision itself: the full registry is never injected into the prompt, and only the selected schema is exposed to the decoder as an output constraint for argument generation.

\paragraph{Efficiency and Context Optimization.}
Recent efforts have focused on memory compression, prompt caching, and long-context models \cite{Kangetal2025}.
These techniques reduce the prompt length but do not change the fact that in-context tool registries scale linearly with the number and verbosity of tools \cite{Wangetal2024}.
NTILC is complementary to long-context and caching techniques: while those methods reduce the cost of what is already in the prompt, NTILC eliminates the registry from the prompt entirely, meaning the two can be composed to reduce both residual prompt overhead and registry footprint simultaneously.

\section{Method}

\paragraph{Problem Setting} Given a user request $\chi$ and a registry of tools $\mathcal{T} = \{t_1, \dots, t_N\}$, the objective is to (i) identify the appropriate tool $t^* \in \mathcal{T}$ that satisfies the request, and (ii) generate a valid set of arguments $a^*$ consistent with the selected tool's interface. Each tool $t \in \mathcal{T}$ is defined by a schema specifying its functionality and input signature. The problem thus requires both semantic alignment between $\chi$ and $t^*$, and functional alignment between $a^*$ and the schema of $t^*$.

\subsection{Inference}
\label{sec:inference}
\begin{algorithm}
\caption{NTILC Tool Loop}
\label{alg:intercept}
\begin{algorithmic}[1]
\Require Query $\chi$, encoder $\mathcal{E}$, tool index $\mathcal{V}$, backbone LLM $\mathcal{M}$, finite state machine $\mathcal{S}$
\State $\mathcal{P} \leftarrow \mathcal{M}(\chi)$ \Comment{LLM produces plan block of tool intents}
\State $\{p_1, \dots, p_k\} \leftarrow \mathrm{Split}(\mathcal{P})$ \Comment{Decompose plan block into individual intents}
\State $v_\chi \leftarrow \mathcal{E}(p_i)$ \Comment{Encode intent to NTILC embedding space}
\State $\hat{t} \leftarrow \mathrm{NNSearch}(\mathcal{V},\, v_\chi)$ \Comment{Retrieve nearest tool cluster from index}
\State $b \leftarrow \mathcal{M}(\chi,\, \mathcal{S}(\hat{t}.\mathrm{schema}))$ \Comment{Constrained argument generation produces dispatch block}
\State $r \leftarrow \mathrm{Dispatch}(\hat{t},\, b)$ \Comment{Execute tool, return response block}
\State $y \leftarrow \mathcal{M}(\chi,\, r)$ \Comment{Synthesize final response}
\State \Return $y$
\end{algorithmic}
\end{algorithm}
\begin{figure}
    \centering
    \includegraphics[width=1\linewidth]{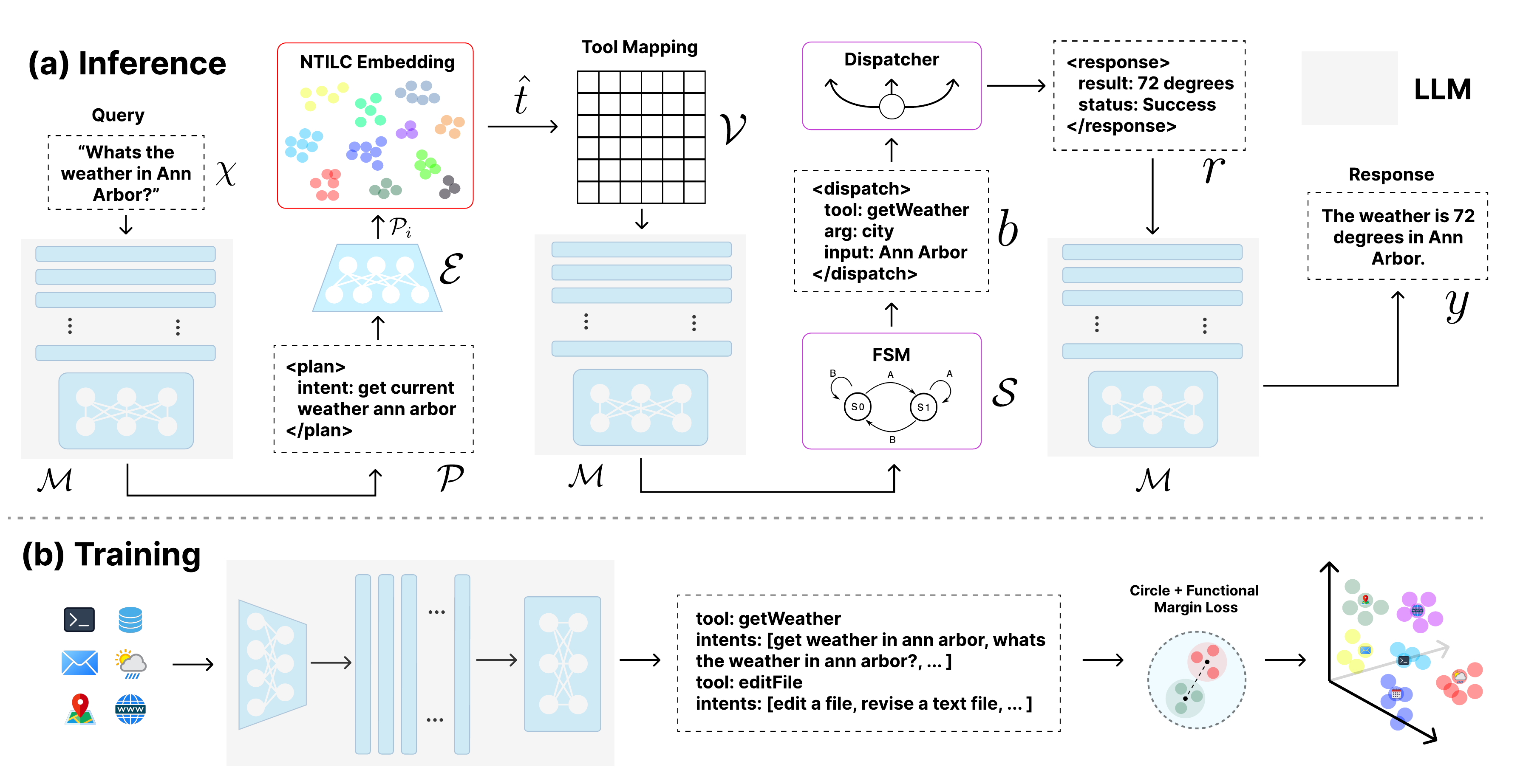}
    \caption{NTILC inference and training pipelines. The backbone LLM $\mathcal{M}$ produces a plan block $\mathcal{P}$ from user request $\chi$, which is decomposed into individual intents and encoded into the NTILC embedding space. Nearest-neighbor search over the pre-built tool index $\mathcal{V}$ retrieves a tool identifier $\hat{t}$, whose schema conditions constrained argument generation via an FSM to produce dispatch block $b$. The dispatcher executes the tool and returns response block $r$ to $\mathcal{M}$, which synthesizes the final response $y$.}
    \label{fig:method}
\end{figure}
As shown in Algorithm~\ref{alg:intercept} and Figure~\ref{fig:method}-(a), at inference time, a backbone LLM $\mathcal{M}$ receives the user request $\chi$ and produces a plan block $\mathcal{P}$ consisting of one or more natural-language tool intents. A splitting function decomposes $\mathcal{P}$ into individual intents $\{p_1, \dots, p_k\}$, each of which is encoded by $\mathcal{E}$ into the trained NTILC embedding space (introduced in Section~\ref{sec:training}) to produce a query vector $v_\chi$. Nearest-neighbor search \cite{johnson2017billionscalesimilaritysearchgpus, douze2025faisslibrary} over the pre-built tool index $\mathcal{V}$ retrieves a cluster identifier $\hat{t}$, which is passed to a mapping function that resolves it to the corresponding tool schema. This schema is handed to the Outlines library \cite{willard2023efficientguidedgenerationlarge}, which enforces a finite-state machine (FSM) constraint over $\mathcal{M}$'s output vocabulary during argument generation, ensuring that the decoder can only produce syntactically valid arguments that conform to the selected tool signature and preventing malformed or schema-incompatible tool calls.
The model then generates a dispatch block $b$ containing syntactically valid arguments for $\hat{t}$. The dispatcher executes the tool call and returns a response block $r$ to $\mathcal{M}$, which synthesizes the final response $y$ for the user.

Because tool embeddings are query-independent, the index $\mathcal{V}$ is built offline and updated incrementally without retraining the encoder. This removes the full tool registry from the active prompt entirely (as shown in Figure~\ref{fig:main}), reducing registry prompt tokens from $O(N)$ to $O(1)$ regardless of how many tools are registered. 

\subsection{Training}
\label{sec:training}

\paragraph{Encoder Architecture.}
The NTILC encoder $\mathcal{E}$ uses a \texttt{all-MiniLM-L6-v2} backbone (22.7M parameters). Final hidden states are mean-pooled and passed through a lightweight two-linear-layer MLP projection head, mapping the pooled representation to a 128-dimensional embedding space. All embeddings are L2-normalized, bounding pairwise distances and stabilizing margin-based objectives across registries. Each tool $t \in \mathcal{T}$ is represented as
\begin{equation}
    t = (n,\ d,\ A,\ R,\ e),
\end{equation}
where $n$ is the name, $d$ is a natural-language description, $A$ is the argument schema, $R$ is the return schema, and $e$ contains metadata. The encoder $\mathcal{E}$ operates on a textual rendering of this schema, while functional signature terms in the loss are derived from $A$, $R$, and $e$.

\paragraph{Training Data.}
To train $\mathcal{E}$, we construct a dataset of $(q, t)$ query–tool pairs, where $q$ is a natural-language intent derived from $\chi$. For each tool in the registry, an LLM generates natural-language intents that a user might express when wanting to invoke that tool, producing a supervised training signal without requiring manual annotation. When dataset-provided train/test splits are available, they are used directly; otherwise, validation splits are stratified from the training data.

\paragraph{Semantic Topology via Circle Loss.}
We adopt Circle Loss \cite{sun2020circlelossunifiedperspective} as the query-tool alignment term. Compared to standard contrastive losses, Circle Loss weights each pairwise similarity score individually, concentrating more gradient signal on examples near the retrieval decision boundary. Let $K^+$ denote the number of positive pairs and $K^-$ denote the number of negative pairs for a given query. The objective is:
\begin{equation}
    \mathcal{L}_{\text{circle}} = \log \left[ 1 + \sum_{j=1}^{K^-} \sum_{i=1}^{K^+} \exp\left(\gamma (\alpha_n^j (s_n^j - \Delta_n) - \alpha_p^i (s_p^i - \Delta_p))\right) \right]
\end{equation}
where $\gamma$ scales the similarity scores and the $\alpha$ factors adaptively weight positive and negative pairs. We set $\gamma = 32.0$, $\Delta_p = 0.75$, $\Delta_n = 0.25$ in all experiments.

\paragraph{Resolving Semantic Blur via Functional Margin Loss.}
Circle Loss alone is insufficient because semantic proximity does not guarantee executable compatibility, as described in Section~\ref{sec:intro}. To handle these cases, we propose the \textbf{Functional Margin (FM) Loss}, which applies an explicit repulsive force to semantic blur hard negatives.

Rather than filtering to a discrete hard-negative set, we assign each negative pair a continuous weight representing the degree of functional incompatibility:
\begin{equation}
    w_{ij} = 1 - \operatorname{compat}(t_i, t_j),
\end{equation}
where $\operatorname{compat}$ is a normalized score in $[0, 1]$ derived from shared argument names, argument types, required fields, and interface metadata. For each tool $t$, define $\operatorname{Sig}(t)$ as the set of required argument names, argument types, return type, and metadata fields. The compatibility score is:
\begin{equation}
    \operatorname{compat}(t_i, t_j) =
    \frac{|\operatorname{Sig}(t_i) \cap \operatorname{Sig}(t_j)|}
    {|\operatorname{Sig}(t_i) \cup \operatorname{Sig}(t_j)|}.
\end{equation}
Pairs whose signatures are nearly identical receive a weight near zero and contribute negligible gradient; pairs with incompatible signatures receive a weight near one and are pushed apart strongly.

Let $z_i, z_j \in \mathbb{R}^d$ be $\ell_2$-normalized tool embeddings. Define the cosine distance $d_{\text{embed}}(t_i, t_j) = 1 - z_i^\top z_j$, a fixed margin hyperparameter $m_0 > 0$, and the margin residual $\rho_{ij} = \max(0,\ m_0 - d_{\text{embed}}(t_i,t_j))$. The FM loss over a batch of $B$ examples is:
\begin{equation}
    \mathcal{L}_{\text{FM}} = \frac{1}{B} \sum_{i=1}^{B}
    \frac{
        \displaystyle\sum_{j \neq g_i} w_{ij}\, \mathbf{1}[\rho_{ij} > 0]\, \rho_{ij}^{2}
    }{
        \displaystyle\sum_{j \neq g_i} w_{ij}\, \mathbf{1}[\rho_{ij} > 0] + \varepsilon
    },
\end{equation}
where $B$ is the batch size, $g_i$ is the index of the ground-truth tool for query $i$, and $\varepsilon$ is a small constant for numerical stability. Only pairs with an active margin violation ($\rho_{ij} > 0$) contribute to the numerator; the denominator normalizes by the total weight of those active pairs rather than the full negative set, concentrating gradient on the hardest functionally-incompatible negatives.

\paragraph{Composite Training Objective.}
The full training objective combines query-tool alignment with signature-aware tool separation:
\begin{equation}
    \mathcal{L}_{\text{dispatch}} = \lambda_1 \mathcal{L}_{\text{circle}} + \lambda_2 \mathcal{L}_{\text{FM}}.
\end{equation}
The $\lambda_1$ coefficient governs the broad semantic structure of the embedding space, while $\lambda_2$ controls the strength of functional discrimination among confusable tools. We sweep $\lambda_1 \in \{0.1, 0.5, 1.0\}$ and $\lambda_2 \in \{0.1, 0.5, 1.0, 2.0, 5.0, 10.0\}$, with additional grids for $m_0$ reported in the appendix. The optimal configuration is $\lambda_1 = 1.0$ and $\lambda_2 = 2.0$.

\section{Experiments}

We evaluate NTILC on two claims: (1) it reduces context length during tool selection, and (2) it preserves tool-retrieval accuracy while reducing inference time.
All results are reported on public tool-selection or function-calling datasets.

All NTILC models were trained and evaluated on a single NVIDIA H100 80GB GPU. Inference latency for NTILC is measured on this local hardware. Training took on average ${\sim}40$ minutes, and inference is performed with batch size 32. For closed-source baselines (ChatGPT, Claude, Gemini), we use their public APIs and do not have access to underlying hardware configurations.

\subsection{Experimental Setup}

We evaluate on five public datasets, each normalized into a common representation:
\begin{equation}
    x = (q, \mathcal{T}, t^*, a^*, \sigma),
\end{equation}
where $q$ is the user request, $\mathcal{T}$ is the available tool registry, $t^*$ is the target tool, $a^*$ contains correct arguments when provided, and $\sigma$ is a dataset evaluator signal.

\begin{table}[H]
\caption{Public dataset coverage. Each row corresponds to an adapter that maps the original dataset into the NTILC tool-selection format. Blur pairs denote tool pairs that are semantically similar but functionally incompatible.}
\label{tab:datasets}
\centering
\small
\begin{tabular}{lccc}
\toprule
Dataset & Primary Signal & Tool Registry Size & Blur Pairs \\
\midrule
ToolBench \cite{qin2023toolllmfacilitatinglargelanguage}    & Pass Rate / Win Rate   & 16,464 & 808 \\
API-Bank \cite{li2023apibankcomprehensivebenchmarktoolaugmented}     & Tool Call Accuracy     & 3,169 & 190 \\
BFCL \cite{patil2025the}         & AST / Exec. Success    & 2,138 & 133 \\
ToolEyes \cite{ye2024tooleyesfinegrainedevaluationtool}     & Planning / Tool Call   & 588 & 49 \\
MetaTool \cite{huang2024metatoolbenchmarklargelanguage}     & Selection / Awareness  & 199 & 32\\
\bottomrule
\end{tabular}
\end{table}

The datasets above allow us to measure how performance and cost scale with registry size.
Semantic-blur evaluation subsets are constructed per dataset by computing $w_{ij} = 1 - \operatorname{compat}(t_i, t_j)$ for all tool pairs and retaining pairs where $w_{ij}$ exceeds a fixed threshold, producing dataset-specific hard-negative subsets used for both training and evaluation.

\subsection{Context Token Cost}

Traditional ICT places the full tool registry in the prompt on every tool-selection call.
Let $S$ be non-tool system-prompt tokens, $Q$ be user-query tokens, $D$ be the fixed NTILC dispatcher instruction, $N$ be the number of available tools, and $\bar{r}$ be the average schema length in tokens.
NTILC keeps the registry outside the prompt as an embedding index, yielding constant prompt length regardless of $N$.

\begin{figure}
    \centering
    \includegraphics[width=1\linewidth]{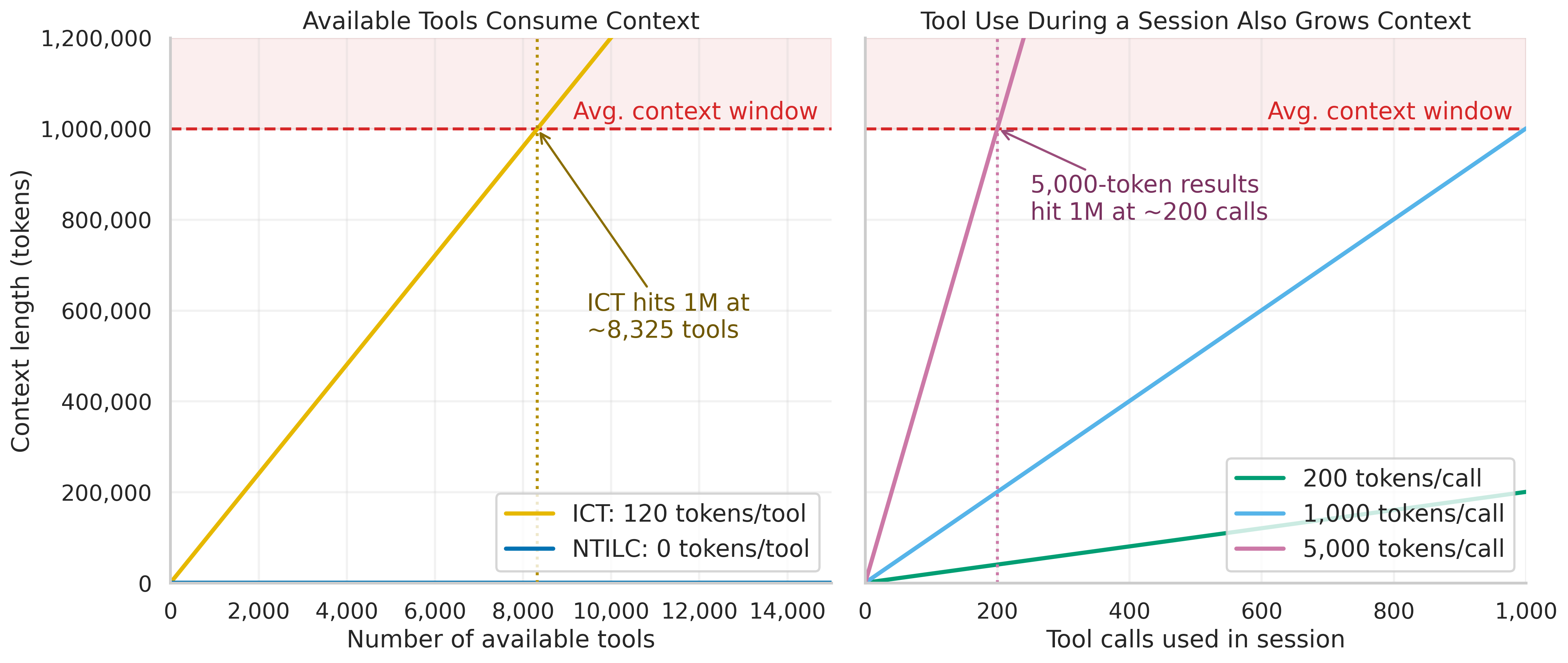}
    \caption{Context length as a function of the number of available tools. ICT incurs linear growth by inserting the full tool registry into the prompt, whereas NTILC externalizes the registry through latent retrieval and preserves an approximately constant context length.}
    \label{fig:context_length_by_tools}
\end{figure}

\begin{table}[H]
\caption{Token cost for ICT versus NTILC. Registry prompt tokens scale linearly with tools under ICT; NTILC keeps the registry external to the LLM prompt.}
\label{tab:congestion}
\centering
\small
\begin{tabular}{lcccc}
\toprule
Method & System Prompt & Registry Prompt Cost & Total Prompt Tokens & Scaling \\
\midrule
ICT & $S$ & $N\bar{r}$ & $S + Q + N\bar{r}$ & $O(N)$ \\
\textbf{NTILC} & $S + D$ & \textbf{0} & $\mathbf{S + D + Q}$ & $\mathbf{O(1)}$ \\
\bottomrule
\end{tabular}
\end{table}

\subsection{Inference Time and Tool-Retrieval Accuracy}

We compare NTILC against ICT baselines in table~\ref{tab:model_comparison} using Qwen3-27B, Ministral 3, Kimi Moonlight, ChatGPT 5, Gemini 2.5 Flash, and Claude Sonnet 4.6.
Each ICT baseline receives the same tool registry in the prompt.
NTILC uses Qwen3-27B for constrained argument generation and the learned tool-embedding index for tool selection.

\begin{table}
\caption{Inference-time performance as tool registry size increases (ICT baseline), evaluated using a Qwen3-27B model on the ToolBench dataset, which enables experiments with large tool registries.}
\label{tab:registry_sweep}
\centering
\small
\begin{tabular}{c c c c c}
\toprule
$\#\text{Tools}$ & Registry Tokens $\downarrow$ & Top-1 $\uparrow$ & Top-5 $\uparrow$ & Latency (ms) $\downarrow$ \\
\midrule
10  & 667   & 1.000 & 1.000 & 4934  \\
50  & 2390  & 0.970 & 1.000 & 5821  \\
100 & 4787  & 0.950 & 0.980 & 6119  \\
150 & 6884  & 0.920 & 0.970 & 7668  \\
200 & 9216  & 0.920 & 0.980 & 7045  \\
250 & 11575 & 0.870 & 0.960 & 12120 \\
\bottomrule
\end{tabular}
\end{table}

\begin{table}
\caption{Main inference-time comparison, including parameter counts. Registry token counts are constant within each dataset, as identical prompts are used across all tools. ``Gen.\ Tokens'' denotes the number of tokens generated by the model during reasoning and tool selection, and latency is reported in milliseconds (ms).}
\label{tab:model_comparison}
\centering
\scriptsize
\setlength{\tabcolsep}{10pt}
\renewcommand{\arraystretch}{0.9}
\begin{adjustbox}{max width=\linewidth}
\begin{tabular}{llrccccc}
\toprule
Dataset & Method & Params & Registry $\downarrow$ & Gen. Tokens $\downarrow$ & Top-1 $\uparrow$ & Top-5 $\uparrow$ & Latency $\downarrow$ \\
\midrule

\multirow{7}{*}{ToolEyes}
& Qwen3-27B (ICT) & 27B & 22757 & 2303 & 93\% & 98\% & 4609 \\
& Ministral 3 & 14B & 22757 & 2518 & 91\% & 97\% & 4892 \\
& Kimi Moonlight & 16B & 22757 & 2864 & 92\% & 98\% & 5127 \\
& ChatGPT 5 (ICT) & 1T+ & 22757 & 3092 & 93\% & 99\% & -- \\
& Gemini 2.5 Flash (ICT) & 1T+ & 22757 & 2580 & 93\% & \textbf{100\%} & -- \\
& Claude Sonnet 4.6 (ICT) & 1T+ & 22757 & 4222 & \textbf{96\%} & \textbf{100\%} & -- \\
& \textbf{NTILC (Ours)} & 27.2B & \textbf{0} & \textbf{115} & 94\% & \textbf{100\%} & \textbf{1127} \\

\midrule
\multirow{7}{*}{MetaTool}
& Qwen3-27B (ICT) & 27B & 6593 & 1911 & \textbf{97\%} & 99\% & 4299 \\
& Ministral 3 & 14B & 6593 & 2048 & 94\% & 98\% & 4516 \\
& Kimi Moonlight & 16B & 6593 & 2375 & 95\% & 99\% & 4868 \\
& ChatGPT 5 (ICT) & 1T+ & 6593 & 2649 & 95\% & \textbf{100\%} & -- \\
& Gemini 2.5 Flash (ICT) & 1T+ & 6593 & 2113 & 92\% & 99\% & -- \\
& Claude Sonnet 4.6 (ICT) & 1T+ & 6593 & 3651 & \textbf{97\%} & \textbf{100\%} & -- \\
& \textbf{NTILC (Ours)} & 27.2B & \textbf{0} & \textbf{162} & \textbf{97\%} & \textbf{100\%} & \textbf{1512} \\

\midrule
\multirow{7}{*}{API-Bank}
& Qwen3-27B (ICT) & 27B & 115189 & 2073 & 95\% & 99\% & 4539 \\
& Ministral 3 & 14B & 115189 & 2294 & 93\% & 98\% & 4811 \\
& Kimi Moonlight & 16B & 115189 & 2667 & 94\% & 99\% & 5194 \\
& ChatGPT 5 (ICT) & 1T+ & 115189 & 2856 & 96\% & 99\% & -- \\
& Gemini 2.5 Flash (ICT) & 1T+ & 115189 & 2236 & 92\% & \textbf{100\%} & -- \\
& Claude Sonnet 4.6 (ICT) & 1T+ & 115189 & 3883 & \textbf{97\%} & \textbf{100\%} & -- \\
& \textbf{NTILC (Ours)} & 27.2B & \textbf{0} & \textbf{137} & 96\% & \textbf{100\%} & \textbf{1268} \\

\midrule
\multirow{7}{*}{BFCL}
& Qwen3-27B (ICT) & 27B & 21288 & 2195 & \textbf{98\%} & 99\% & 4725 \\
& Ministral 3 & 14B & 21288 & 2391 & 95\% & 98\% & 5018 \\
& Kimi Moonlight & 16B & 21288 & 2716 & 96\% & 99\% & 5332 \\
& ChatGPT 5 (ICT) & 1T+ & 21288 & 2983 & 94\% & 99\% & -- \\
& Gemini 2.5 Flash (ICT) & 1T+ & 21288 & 2365 & 92\% & 99\% & -- \\
& Claude Sonnet 4.6 (ICT) & 1T+ & 21288 & 4074 & 97\% & 99\% & -- \\
& \textbf{NTILC (Ours)} & 27.2B & \textbf{0} & \textbf{153} & \textbf{98\%} & \textbf{100\%} & \textbf{1398} \\

\midrule
\multirow{7}{*}{ToolBench}
& Qwen3-27B (ICT) & 27B & 386355 & 2446 & 98\% & 98\% & 5663 \\
& Ministral 3 & 14B & 386355 & 2685 & 96\% & 98\% & 5924 \\
& Kimi Moonlight & 16B & 386355 & 3217 & 97\% & \textbf{99\%} & 6410 \\
& ChatGPT 5 (ICT) & 1T+ & 386355 & 3688 & 99\% & \textbf{100\%} & -- \\
& Gemini 2.5 Flash (ICT) & 1T+ & 386355 & 2788 & 95\% & 99\% & -- \\
& Claude Sonnet 4.6 (ICT) & 1T+ & 386355 & 5094 & 97\% & 99\% & -- \\
& \textbf{NTILC (Ours)} & 27.2B & \textbf{0} & \textbf{147} & 98\% & \textbf{100\%} & \textbf{1452} \\

\bottomrule
\end{tabular}
\end{adjustbox}
\end{table}

\begin{figure}
    \centering
    \includegraphics[width=1\linewidth]{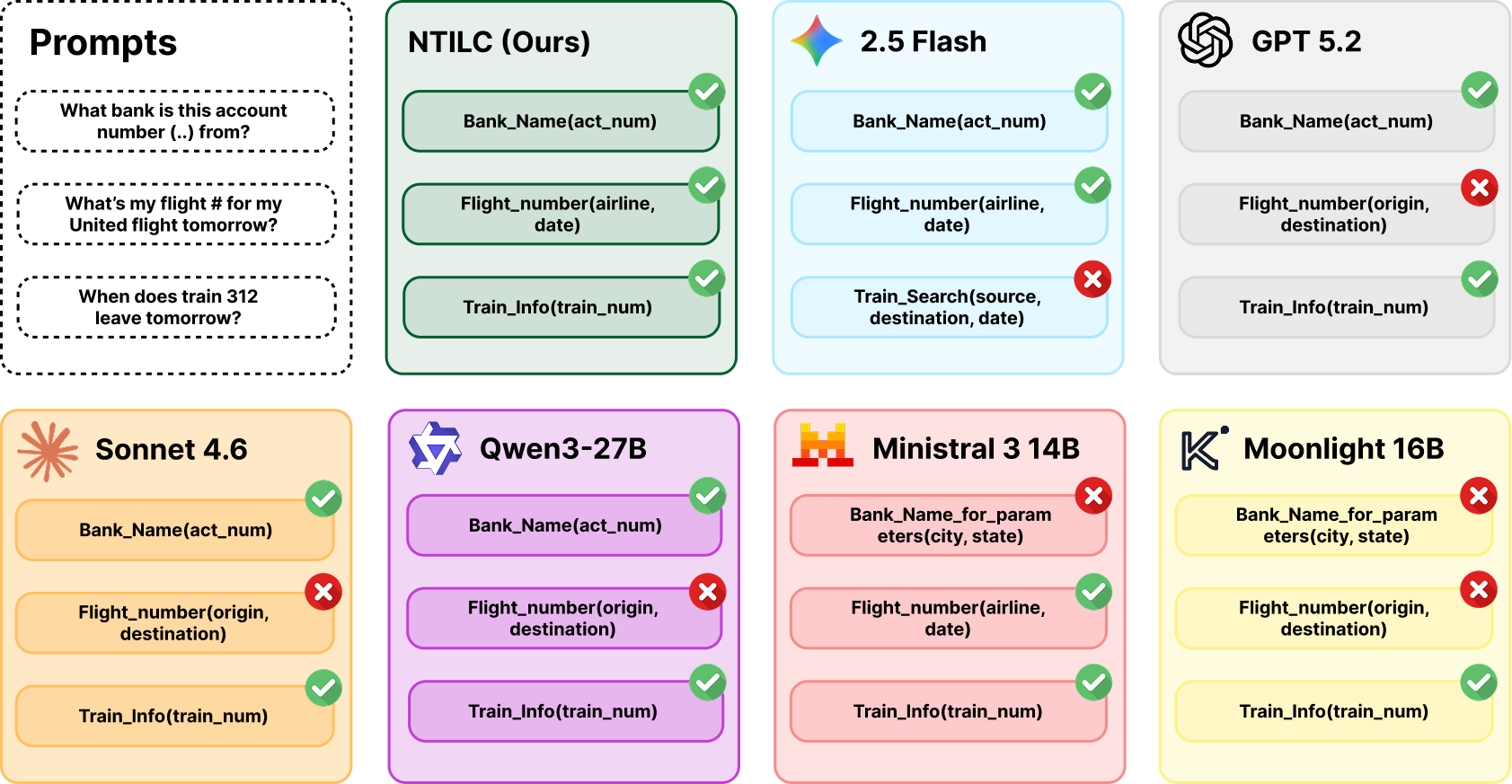}
    \caption{Qualitative examples illustrating \textit{semantic blur}. While baseline methods frequently misclassify user intents by selecting tools with similar natural language descriptions but executionally incompatible arguments, NTILC leverages Functional Margin (FM) Loss to separate these confusable tools in the embedding space, ensuring accurate and functionally viable tool selection.}
    \label{fig:qualitative}
\end{figure}

\subsection{Retrieval and Loss Ablations}
\label{sec:ablation}

Table~\ref{tab:retrieval} isolates the tool-retrieval component to evaluate different encoders and training objectives. This table provides the primary evidence that our proposed Functional Margin Loss ($\mathcal{L}_{\text{FM}}$) is the specific driver for performance improvements under semantic blur. While moving from sparse retrieval (BM25) to a strong off-the-shelf dense encoder (Qwen3-Embedding-8B) provides a baseline performance jump, the critical ablation lies in the bottom two rows. When the embeddings are fine-tuned using standard metric learning (Circle Loss only), the Semantic-Blur Accuracy reaches 70.8\%. However, by augmenting the objective with our Functional Margin Loss (NTILC), the Semantic-Blur Accuracy increases significantly to 75.0\%, alongside a drop in Functional Error to 8.7\%. Because the underlying encoder architecture remains identical between these two configurations, this 4.2\% absolute improvement clearly demonstrates that $\mathcal{L}_{\text{FM}}$, and not just the encoder itself, is directly responsible for teaching the model to resolve ambiguous, semantically blurred queries.

\begin{table}
\caption{Tool-retrieval ablation on pooled public dataset test splits. Functional error rate measures the fraction of selected tools whose specification is functionally incompatible with the correct tool.}
\label{tab:retrieval}
\centering
\small
\begin{tabular}{lcccc}
\toprule
Retriever / Loss & Top-1 Acc. $\uparrow$ & Top-5 Acc. $\uparrow$ & Semantic-Blur Acc. $\uparrow$ & Functional Error $\downarrow$ \\
\midrule
BM25 & 26.6\% & 44.7\% & 12.5\% & 72.7\% \\
Qwen3-Embedding-8B & 82.4\% & 95.4\% & 66.7\% & 17.1\% \\
Circle Loss only & 89.9\% & 96.4\% & 70.8\% & 10.1\% \\
\textbf{NTILC ($\mathcal{L}_{\text{FM}}$)} & \textbf{91.3\%} & \textbf{97.4\%} & \textbf{75.0\%} & \textbf{8.7\%} \\
\bottomrule
\end{tabular}
\end{table}

\section{Conclusion}

NTILC achieves substantial reductions in prompt token cost and selection latency without sacrificing accuracy.
On ToolBench, it reaches 98\% Top-1 accuracy with zero registry prompt tokens, maintaining parity with state-of-the-art ICT models while reducing selection latency by approximately 74\% (from 5663ms to 1452ms).
The Functional Margin Loss is the key driver of gains on semantic-blur subsets, as confirmed by ablations.
More broadly, NTILC demonstrates that the tool registry need not live in the context window: moving it to a compact embedding index decouples agent capability from prompt length, a property that becomes increasingly important as tool registries grow.

\paragraph{Limitations and Future Work.}
NTILC requires the tool registry to be indexed ahead of time.
Static tool schemas are easily handled, but tools with highly dynamic or state-dependent documentation may require frequent re-indexing or a hybrid approach that retrieves a short schema snippet at query time.
NTILC also focuses on selecting the correct tool; argument generation and tool-execution safety remain separate evaluation surfaces.
Future work will examine larger and more dynamic registries, online index updates when tools are added or modified, and richer functional-distance metrics that capture schema compatibility beyond argument names and primitive types.

\paragraph{Broader Impact.}
NTILC changes how tools are selected, not what policy governs their use.
A malicious or ambiguous query could still cause a model to invoke an unintended tool, particularly when the registry contains tools with external side effects.
We recommend pairing NTILC with allow-lists, per-tool risk labels, and pre-dispatch policy filters.
The efficiency gains also lower the barrier to deploying large-registry agents, which warrants care around intent-level safety filtering applied before the dispatch step.

\bibliographystyle{abbrvnat}
\bibliography{references}

% \newpage

% \include{completed_checklist}

\end{document}